\documentclass{article}
\usepackage{amsmath}
\usepackage{amssymb}
\usepackage{makeidx}
\usepackage[dvips]{graphicx}

\begin{document}

\title{{\huge Metric gauge fields in Deformed Special Relativity }}
\author{{\large Roberto Mignani}$^{1-3},${\large Fabio Cardone}$^{2,4}$%
{\large \ \ and Andrea Petrucci}$^{5}$ {\large \ } \\
$^{1}$Dipartimento di Fisica \textquotedblright E.Amaldi\textquotedblright ,
Universit\`{a} degli Studi \textquotedblright Roma Tre\textquotedblright \\
\ Via della Vasca Navale, 84 - 00146 Roma, Italy\\
$^{2}$GNFM, Istituto Nazionale di Alta Matematica "F.Severi"\\
\ Citt\`{a} Universitaria, P.le A.Moro 2 - 00185 Roma, Italy\\
$^{3}$I.N.F.N. - Sezione di Roma III\\
$^{4}$Istituto per lo Studio dei Materiali Nanostrutturati (ISMN -- CNR)\\
Via dei Taurini - 00185 Roma, Italy\\
$^{5}$ENEA, Italian National Agency for new Technologies, \\Energy
and
sustainable economic development \\
Via Anguillarese 301 - 00123 Roma, Italy \\
}
\maketitle
\date{}

\begin{abstract}
We show that, in the framework of Deformed Special Relativity\emph{\ }(DSR),
namely a (four-dimensional) generalization of the (local) space-time
structure based on an energy-dependent \textquotedblright
deformation\textquotedblright\ of the usual Minkowski geometry, two kinds of
gauge symmetries arise, whose spaces either coincide with the deformed
Minkowski space or are just internal spaces to it. This is why we named them
\emph{"metric gauge theories"}. In the case of the internal gauge fields,
they are a consequence of the deformed Minkowski space (DMS) possessing the
structure of a generalized Lagrange space. Such a geometrical structure
allows one to define curvature and torsion in the DMS.
\end{abstract}

\newpage

\section{Introduction}

It is well known that gauge theories play presently a basic role in
describing all the known interactions. In all cases, gauge symmetries are
related to physical fields directly arising from the symmetries ruling some
given interaction; on one side, this leads to the rising of a new, dynamical
gauge field; on the other hand, if the gauge symmetry is broken, such a
circumstance provides one with new --- often unforeseen --- informations
about the structural properties of the interaction considered.

Often, as known as well, in spite of the fact that the physical world is the
usual Minkowski space-time, the gauge manifold is \textit{not} the usual,
Minkowski one. For instance, in the case of the usual Minkowski space, the
gauge symmetry of electrodynamics does actually work in an auxiliary space
(Weyl charge space). It is therefore worth to investigate when and where
gauge symmetries can be introduced in a Minkowski space, and to lead to
significant physical results.

It is just the purpose of the present paper to show that \ this circumstance
occurs in the framework of \emph{Deformed Special Relativity (DSR),} namely
a (four-dimensional) generalization of the (local) space-time structure
based on an energy-dependent \textquotedblright
deformation\textquotedblright\ of the usual Minkowski geometry \cite%
{carmig1,carmig2}. As we shall see, in DSR two kinds of gauge symmetries
arise, whose spaces either coincide with the deformed Minkowski space (DMS) $%
\widetilde{M}$ or are just internal spaces to it. This is why we named them
\emph{"metric gauge theories".}

The paper is organized as follows. In Sect.2, we review the basic features
of DSR that are relevant to our purposes. Sect.3 discuss DSR as a metric
gauge theory. Metric gauge fields can be external (Subsect.3.1) or internal
(Subsect.3.2). The last topic is related to the structure of DMS as
generalized Lagrange space, whose main properties are summarized. Subsect.
3.2.2 deals with the structure of $\widetilde{M}$ as generalized Lagrange
space. The internal gauge fields of $\widetilde{M}$ are discussed in
Subsect.3.2.4. In 3.3 we present a possible experimental evidence for such
metric gauge fields. Conclusions and perspectives are given in Sect.4.

\bigskip

\section{Elements of Deformed Special Relativity}

\subsection{Energy and Geometry}

The geometrical structure of the physical world --- both at a large and a
small scale \ --- has been debated since a long. After Einstein, the
generally accepted view considers the arena of physical phenomena as a
four-dimensional spacetime, endowed with a \emph{global}\textit{, }curved,
Riemannian structure and a \emph{local}\textit{, }flat, Minkowskian geometry.

However, an analysis of some experimental data concerning physical phenomena
ruled by different fundamental interactions have provided evidence for a
local departure from Minkowski metric \cite{carmig1,carmig2}: among them,
the lifetime of the (weakly decaying) \textit{K$_{s}^{0}$} meson, the
Bose-Einstein correlation in (strong) pion production and the superluminal
propagation of electromagnetic waves in waveguides. These phenomena
seemingly show a (local) breakdown of Lorentz invariance, together with a
plausible inadequacy of the Minkowski metric; on the other hand, they can be
interpreted in terms of a deformed Minkowski spacetime, with metric\textit{\
}coefficients depending on the energy of the process considered \cite%
{carmig1,carmig2}.

All the above facts suggested to introduce a (four-dimensional)
generalization of the (local) space-time structure based on an
energy-dependent \textquotedblright deformation\textquotedblright\ of the
usual Minkowski geometry of $M$, whereby the corresponding deformed metrics
ensuing from the fit to the experimental data seem to provide an \emph{%
effective dynamical description of the relevant interactions}\textit{\ } (%
\emph{at the energy scale and in the energy range considered}).

An analogous energy-dependent metric seems to hold for the gravitational
field (at least locally, i.e. in a neighborhood of Earth) when analyzing
some classical experimental data concerning the slowing down of clocks.

Let us shortly review the main ideas and results concerning the
(four-dimensional ) deformed Minkowski spacetime $\widetilde{M}$.

The four-dimensional \textquotedblright deformed\textquotedblright\ metric
scheme is based on the assumption that spacetime, in a preferred frame which
is \emph{fixed} by the scale of energy $E$ , is endowed with a metric of the
form

\begin{equation}
\begin{array}{c}
ds^{2}=b_{0}^{2}(E)c^{2}dt^{2}-b_{1}^{2}(E)dx^{2}-b_{2}^{2}(E)dy^{2}-b_{3}^{2}(E)dz^{2}=g_{DSR\mu \nu }(E)dx^{\mu }dx^{\nu };
\\
g_{DSR\mu \nu }(E)=\left(
b_{0}^{2}(E),-b_{1}^{2}(E),-b_{2}^{2}(E),-b_{3}^{2}(E)\right) ,%
\end{array}
\label{1}
\end{equation}%
with $x^{\mu }=(x^{0},x^{1},x^{2},x^{3})=(ct,x,y,z),$ $c$ being the usual
speed of light in vacuum. We named \textquotedblright Deformed Special
Relativity\textquotedblright\ (DSR) the relativity theory built up on metric
(1).

Metric (1) is supposed to hold locally, i.e. in the spacetime region where
the process occurs. It is supposed moreover to play a \emph{dynamical} role,
and to provide a geometric description of the interaction considered. In
this sense, DSR realizes the so called \emph{"Finzi Principle of Solidarity"
}between space-time and phenomena occurring in it\emph{\ }\footnote{%
Let us recall that in 1955 the Italian mathematician Bruno Finzi \textit{\ }%
stated his \emph{\textquotedblright Principle of
Solidarity\textquotedblright }(PS), that sounds \textquotedblright \textit{%
It's (indeed) necessary to consider space-time TO\ BE SOLIDLY\ CONNECTED
with the physical phenomena occurring in it, so that its features and its
very nature do change with the features and the nature of those. In this way
not only (as in classical and special-relativistic physics)\ space-time
properties affect phenomena, but reciprocally phenomena do affect space-time
properties. One thus recognizes in such an appealing \textquotedblright
Principle of Solidarity\textquotedblright\ between phenomena and space-time
that characteristic of mutual dependence between entities, which is peculiar
to modern science.}\textquotedblright\ Moreover, referring to a generic
N-dimensional space: \textit{\textquotedblright\ It can, }a priori\textit{,
be pseudoeuclidean, Riemannian, non-Riemannian. But} --- he wonders ---
\textit{how is indeed the space-time where physical phenomena take place?
Pseudoeuclidean, Riemannian, non-Riemannian, according to their nature, as
requested by the principle of solidarity between space-time and phenomena
occurring in it.\textquotedblright\ }
\par
Of course, Finzi's main purpose was to apply such a principle to Einstein's
Theory of General Relativity, namely to the class of gravitational
phenomena. However, its formulation is as general as possible, so to apply
in principle to all the known physical interactions. Therefore, Finzi's PS
is at the very ground of any attempt at geometrizing physics, i.e.
describing physical forces in terms of the geometrical structure of
space-time.}(see \cite{finzi}). Futhermore, we stress that, from the
physical point of view, $E$\emph{\ is the measured energy of the system},
and thus a merely phenomenological (non-metric) variable\footnote{%
As is well known, all the present physically realizable detectors work
\textit{via} their electromagnetic interaction in the usual space-time $M$.
So, $E$ is the energy of the system measured in \emph{fully Minkowskian
conditions.}}.

We notice explicitly that the spacetime $\widetilde{M}$ described by (1) is
flat (it has zero four-dimensional curvature), so that the geometrical
description of the fundamental interactions based on it differs from the
general relativistic one (whence the name \textquotedblright
deformation\textquotedblright\ used to characterize such a situation).
Although for each interaction the corresponding metric reduces to the
Minkowskian one for a suitable value of the energy $E_{0}$ (which is
characteristic of the interaction considered), the energy of the process is
fixed and cannot be changed at will. Thus, in spite of the fact that \emph{%
formally} it would be possible to recover\ the usual Minkowski space $M$ by
a suitable change of coordinates (e.g. by a rescaling), this would amount,
in such a framework, to be a mere mathematical operation devoid of any
physical meaning.

As far as phenomenology is concerned, it is important to recall that a local
breakdown of Lorentz invariance may be envisaged for all the four
fundamental interactions (electromagnetic, weak, strong and gravitational)
whereby\textit{\ }\emph{one gets evidence for a departure of the spacetime
metric from the Minkowskian one} ( in the energy range examined). The
explicit functional form of the metric (1) for all the four interactions can
be found in \cite{carmig1,carmig2}. Here, we confine ourselves to recall the
following basic features of these energy-dependent phenomenological metrics:

1) Both the electromagnetic and the weak metric show the same functional
behavior, namely%
\begin{equation}
g_{DSR\mu \nu }(E)=diag\left( 1,-b^{2}(E),-b^{2}(E),-b^{2}(E)\right) ;
\label{2}
\end{equation}%
\begin{equation}
b^{2}(E)=\left\{
\begin{array}{llll}
(E/E_{0})^{1/3}, & 0\leq E\leq E_{0} &  &  \\
1, & E_{0}\leq E &  &
\end{array}%
\right.  \label{3}
\end{equation}%
with the only difference between them being the threshold energy $E_{0}$ ,
i.e. the energy value at which the metric parameters are constant, i.e. the
metric becomes Minkowskian; the fits to the experimental data yield
\begin{equation}
E_{0,e.m.}=5.0\pm 0.2\mu eV\text{ ; }E_{0w}=80.4\pm 0.2GeV;  \label{4}
\end{equation}

2) for strong and gravitational interactions, the metrics read:
\begin{equation}
g_{DSR}(E)=diag\left(
b_{0}^{2}(E),-b_{1}^{2}(E),-b_{2}^{2}(E),-b_{3}^{2}(E)\right) ;  \label{5}
\end{equation}%
\begin{eqnarray}
b_{0,strong}^{2}(E) &=&b_{3,strong}^{2}(E)=\left\{
\begin{array}{ll}
1, & 0\leq E<E_{0strong} \\
(E/E_{0strong})^{2}, & E_{0strong}<E%
\end{array}%
\right. ;  \notag \\
b_{1,strong}^{2}(E) &=&\left( \sqrt{2}/5\right)
^{2};b_{2,strong}^{2}=(2/5)^{2};  \label{6}
\end{eqnarray}

\begin{equation}
b_{0,grav}^{2}(E)=\left\{
\begin{array}{ll}
1\text{ }, & 0\leq E<E_{0grav} \\
\frac{1}{4}(1+E/E_{0grav})^{2}, & E_{0grav}<E%
\end{array}%
\right.  \tag{6'}
\end{equation}%
with
\begin{equation}
E_{0s}=367.5\pm 0.4GeV;E_{0grav}=20.2\pm 0.1\mu eV.  \label{7}
\end{equation}%
Let us stress that, in this case, contrarily to the electromagnetic and the
weak ones, \emph{a deformation of the time coordinate occurs; }moreover,
\emph{the three-space is anisotropic\footnote{%
At least for strong interaction; nothing can be said for the gravitational
one.}}, with two spatial parameters constant (but different in value) and
the third one variable with energy in an \emph{\textquotedblright
over-Minkowskian\textquotedblright } way (namely it reaches the limit of
Minkowskian metric for decreasing values of $E$ , with $E>E_{0}$) \cite%
{carmig1,carmig2}.

As a final remark, we stress that actually \emph{the four-dimensional
energy-dependent spacetime\ }$\widetilde{M}$ \emph{is just a manifestation
of a larger, five-dimensional space in which energy plays the role of a
fifth dimension.} Indeed, it can be shown that the physics of the
interaction lies in the curvature of such a five-dimensional spacetime, in
which the four-dimensional, deformed Minkowski space is embedded. Moreover,
\emph{all} the phenomenological metrics (2), (3) and (5), (6) can be
obtained as solutions of the vacuum Einstein equations in this generalized
Kaluza-Klein scheme \cite{carmig1,carmig2}.

\subsection{Field Deformation}

We want now to show that the deformation of space-time, expressed by the
metric $g_{DSR}$ (Eq.(1)), does affect also the external fields applied to
the physical system considered.

Let us consider for instance the case of a physical process ruled by the
electromagnetic interaction. Therefore, the Minkowski space $M$ is endowed
with the electromagnetic tensor $F_{\mu \nu }(x)$ (external e.m. field)
acting on the system. Of course $F_{\nu }^{\mu }(x)=g_{SR}^{\mu \rho
}F_{\rho \nu }(x)$.

In the deformed Minkowski space $\widetilde{M}$, the covariant components of
the electromagnetic tensor read
\begin{equation}
\widetilde{F}_{\mu \nu }=g_{DSR\mu \rho }F_{\nu }^{\rho }=g_{DSR\mu \rho
}{}g_{SR}^{\mu \sigma }F_{\sigma \nu },  \label{8}
\end{equation}%
where
\begin{equation}
(g_{DSR\mu \rho }{}g_{SR}^{\mu \sigma })=diag(b_{0}^{2},b_{1}^{2},b\text{$%
_{2}^{2},b_{3}^{2})=($}b_{\sigma }^{2}\delta _{\rho }^{\sigma }).  \label{9}
\end{equation}%
We have therefore
\begin{equation}
\widetilde{F}_{0\nu }=b_{0}^{2}F_{0\nu };\widetilde{F}_{1\nu
}=b_{1}^{2}F_{1\nu };\widetilde{F}_{2\nu }=b_{2}^{2}F_{2\nu };\widetilde{F}%
_{3\nu }=b_{3}^{2}F_{3\nu },  \label{10}
\end{equation}%
or
\begin{equation}
\widetilde{F}_{\mu \nu }=b_{\mu }^{2}F_{\mu \nu },\text{ \ \ \ \ \ \ \ \ \ }%
\mu ,\nu =0,1,2,3  \label{11}
\end{equation}%
(no sum on repeated indices!).

It follows that the tensor $\widetilde{F}_{\mu \nu }$ is not antisymmetric:
\begin{equation}
\widetilde{F}_{\mu \nu }\neq -\widetilde{F}_{\nu \mu }.  \label{12}
\end{equation}

The result shown here for the electromagnetic interaction can be generalized
to other fundamental interactions described by tensor fields.

On account of the well-known identification
\begin{equation}
\widetilde{F}_{0i}=\widetilde{E_{i}},\widetilde{F}_{12}=-\widetilde{B_{3}},%
\widetilde{F}_{23}=-\widetilde{B_{1}},\widetilde{F}_{31}=-\widetilde{B_{2}}
\label{13}
\end{equation}%
(and analogously for $F_{\mu \nu }$), we can write, for the energy density $%
\widetilde{\mathcal{E}}$ of the deformed electromagnetic field:
\begin{equation}
\widetilde{\mathcal{E}}=\frac{\widetilde{\mathbf{E}}^{2}+\widetilde{\mathbf{B%
}}^{2}}{8\pi }=\frac{b_{0}^{4}\mathbf{E}%
^{2}+b_{1}^{4}B_{3}^{2}+b_{2}^{4}B_{1}^{2}+b_{3}^{4}B_{2}^{2}}{8\pi },
\label{14}
\end{equation}%
to be compared with the standard expression for the e.m. field $\mathbf{E}$
, $\mathbf{B}$:
\begin{equation}
\mathcal{E}=\frac{\mathbf{E}^{2}+\mathbf{B}^{2}}{8\pi }.  \label{15}
\end{equation}

\emph{There is therefore a difference in the energy associated to the
electromagnetic field in the deformed space-time region.} We have, for the
energy density
\begin{equation}
\Delta \mathcal{E}=\mathcal{E}-\widetilde{\mathcal{E}}.  \label{16}
\end{equation}%
We can state that the difference $\Delta \mathcal{E}$ represents \emph{the
energy spent by the interaction in order to deform the space-time geometry.}

We can therefore conclude that \emph{the deformation of space-time does
affect the field itself that deforms the geometry of the space}. There is
therefore a feedback between space and interaction which fully implements
the Solidarity Principle.

\section{DSR as Metric Gauge Theory}

\subsection{External metric gauge fields}

It is clear from the discussion of the phenomenological metrics describing
the four fundamental interactions in DSR that the Minkowski space $M$ is the
space-time manifold of background of any experimental measurement and
detection (namely, of any process of acquisition of information on physical
reality). In particular, we can consider this Minkowski space as that
associated to the electromagnetic interaction above the threshold energy $%
E_{0,e.m.}$. Therefore, in modeling the physical phenomena, one has to take
into account this fact. The geometrical nature of interactions, \textit{i.e.}
assuming the validity of the Finzi principle, means that one has to suitably
\emph{gauge} (with reference to $M$) the space-time metrics with respect to
the interaction \ --- and/or the phenomenon --- under study. In other words,
one needs to \textquotedblright adjust\textquotedblright\ suitably the local
metric of space-time according to the interaction acting in the region
considered. We can name such a procedure \emph{\textquotedblright Metric
Gaugement Process\textquotedblright } (M.G.P.). Like in usual gauge theories
a different phase is chosen in different space-time \emph{points}, in DSR
different metrics are associated to different space-time \emph{manifolds}
according to the interaction acting therein. We have thus a gauge structure
on the space of manifolds
\begin{equation}
\widetilde{\mathcal{M}}\equiv \cup _{g_{DSR}\in \mathcal{P}(E)}\widetilde{M}%
\left( g_{DSR}\right) ,  \label{17}
\end{equation}%
where $\mathcal{P}(E)$ is the set of the energy-dependent pseudoeuclidean
metrics of the type (1). This is why it is possible to regard Deformed
Special Relativity as a \emph{Metric Gauge Theory}. In this case, we can
consider the related fields as \emph{external metric gauge fields.}

However, let us notice that DSR can be considered as a metric gauge theory
from another point of view, on account of the dependence of the metric
coefficients on the energy. Actually, once the MGP has been applied, by
selecting the suitable gauge (namely, the suitable \emph{functional form} of
the metric) according to the interaction considered (thus implementing the
Finzi principle), the metric dependence on the energy implies another
different gauge process. Namely, the metric is gauged according to the
process under study, thus selecting the \emph{given} metric, with the\emph{\
given} \emph{values} of the coefficients, suitable for the given phenomenon.

We have therefore a \emph{double} metric gaugement, according, on one side,
to the interaction ruling the physical phenomenon examined, and on the other
side to its energy, in which the metric coefficients are the analogous of
the gauge functions\footnote{%
The analogy of this second kind of metric gauge with the standard,
non-abelian gauge theories is more evident in the framework of the
five-dimensional space-time $\Re _{5}$ (with energy as extra dimension)
embedding $\widetilde{M}$, on which Deformed Relativity in Five Dimensions
(DR5) is based (see \cite{carmig1,carmig2}). In $\Re _{5}$, in fact, energy
is no longer a parametric variable, like in DSR, but plays the role of fifth
(metric) coordinate. The invariance under such a metric gauge, not manifest
in four dimensions, is instead recovered in the form of the isometries of
the five-dimensional space-time-energy manifold $\Re _{5}$.} .

\subsection{Internal Metric Gauge Fields}

We want now to show that the deformed Minkowski space $\widetilde{M}$ of
Deformed Special Relativity does possess another well-defined geometrical
structure, besides the deformed metrical one. Precisely, we will show that $%
\widetilde{M}$ \ is a \emph{generalized Lagrange space} \cite{jan}. As we
shall see, this implies that DSR admits a different, \emph{intrinsic }gauge
structure.

\subsubsection{\textbf{Deformed Minkowski Space as Generalized Lagrange Space%
}}

\paragraph{Generalized Lagrange Spaces}

Let us give the definition of generalized Lagrange space \cite{miron}, since
usually one is not acquainted with it.

Consider a N-dimensional, differentiable manifold $\mathcal{M}$ and its
(N-dimensional) tangent space in a point, $T\mathcal{M}_{\mathbf{x}}$ ($%
\mathbf{x}\in \mathcal{M}$). As is well known, the union
\begin{equation}
\bigcup_{\mathbf{x}\in \mathcal{M}}T\mathcal{M}_{\mathbf{x}}\equiv T\mathcal{%
M}  \label{18}
\end{equation}%
has a fibre bundle structure. Let us denote by $\mathbf{y}$ the generic
element of $T\mathcal{M}_{\mathbf{x}}$, namely a vector tangent to $\mathcal{%
M}$ in $\mathbf{x}$. Then, an element $u\in T\mathcal{M}$ is a vector
tangent to the manifold in some point $\mathbf{x}\in \mathcal{M}$. Local
coordinates for $T\mathcal{M}$ are introduced by considering a local
coordinate system $(x^{1},x^{2},...,x^{N}$) on $\mathcal{M}$ and the
components of $y$ in such a coordinate system $(y^{1},y^{2},...,y^{N})$. The
$2N$ numbers $(x^{1},x^{2},...,x^{N}$,$y^{1},y^{2},...,y^{N})$ constitute a
local coordinate system on $T\mathcal{M}$. We can write synthetically $u=(%
\mathbf{x},\mathbf{y})$. $T\mathcal{M}$ is a $2N$-dimensional,
differentiable manifold.

Let $\pi $ be the mapping (\emph{natural projection}) $\pi :$ $u=(\mathbf{x},%
\mathbf{y})\longrightarrow \mathbf{x}$. ($\mathbf{x}\in \mathcal{M}$, $%
\mathbf{y\in }T\mathcal{M}_{\mathbf{x}}$). Then, the tern $(T\mathcal{M},\pi
,$ $\mathcal{M}$) is the \emph{tangent bundle} to the base manifold $%
\mathcal{M}$. The image of the inverse mapping $\pi ^{-1}(\mathbf{x})$ is of
course the tangent space $T\mathcal{M}_{\mathbf{x}}$, which is called the
\emph{fiber corresponding to the point }$\mathbf{x}$\emph{\ in the fiber
bundle} \ One considers also sometimes the manifold $\widehat{T\mathcal{M}}=T%
\mathcal{M}/\left\{ 0\right\} $, where $0$ is the zero section of the
projection $\pi $. We do not dwell further on the theory of the fiber
bundles, and refer the reader to the wide and excellent literature on the
subject \cite{sten}.

The natural basis of the tangent space $T_{u}(T\mathcal{M})$ at a point $u=(%
\mathbf{x},\mathbf{y})\in T\mathcal{M}$ is $\left\{ \dfrac{\partial }{%
\partial x^{i}},\dfrac{\partial }{\partial y^{j}}\right\} $, $i,j=1,2,...,N$.

A local coordinate transformation in the differentiable manifold $T\mathcal{M%
}$ reads
\begin{equation}
\left\{
\begin{array}{c}
x^{\prime i}=x^{\prime i}(\mathbf{x}),\text{ \ \ \ }\det \left( \dfrac{%
\partial x^{\prime i}}{\partial x^{j}}\right) \neq 0, \\
y^{\prime i}=\dfrac{\partial x^{\prime i}}{\partial x^{j}}y^{j}.%
\end{array}%
\right.  \label{19}
\end{equation}%
Here, $y^{i}$ is \emph{the Liouville vector field} on $T\mathcal{M}$, i.e. $%
y^{i}\dfrac{\partial }{\partial y^{i}}$.

On account of Eq.(19), the natural basis of $T\mathcal{M}_{\mathbf{x}}$ can
be written as
\begin{equation}
\left\{
\begin{array}{c}
\dfrac{\partial }{\partial x^{i}}=\dfrac{\partial x^{\prime k}}{\partial
x^{i}}\dfrac{\partial }{\partial x^{\prime k}}+\dfrac{\partial y^{\prime k}}{%
\partial x^{i}}\dfrac{\partial }{\partial y^{\prime k}},\bigskip \\
\dfrac{\partial }{\partial y^{j}}=\dfrac{\partial y^{\prime k}}{\partial
y^{j}}\dfrac{\partial }{\partial y^{\prime k}}.%
\end{array}%
\right.  \label{20}
\end{equation}%
Second Eq.(20) shows therefore that the vector basis $\left( \dfrac{\partial
}{\partial y^{j}}\right) $, $j=1,2,...,N$, generates a distribution $%
\mathcal{V}$ defined everywhere on $T\mathcal{M}$ and integrable, too (\emph{%
vertical distribution on} $T\mathcal{M}$).

If $\mathcal{H}$ is a distribution on $T\mathcal{M}$ supplementary to $%
\mathcal{V}$, namely
\begin{equation}
T_{u}(T\mathcal{M})=\mathcal{H}_{u}\oplus \mathcal{V}_{u}\text{ \ \ , \ }%
\forall u\in T\mathcal{M,}  \label{21}
\end{equation}%
then $\mathcal{H}$ is called a \emph{horizontal distribution}, or a \emph{%
nonlinear connection} on $T\mathcal{M}$. A basis for the distributions $%
\mathcal{H}$ and $\mathcal{V}$ are given respectively by $\left\{ \dfrac{%
\delta }{\delta x^{i}}\right\} $ and $\left\{ \dfrac{\partial }{\partial
y^{j}}\right\} $, where the basis in $\mathcal{H}$ explicitly reads
\begin{equation}
\dfrac{\delta }{\delta x^{i}}=\frac{\partial }{\partial x^{i}}-H_{\text{ }%
i}^{j}(\mathbf{x},\mathbf{y})\frac{\partial }{\partial y^{j}}.  \label{22}
\end{equation}%
Here, $H_{\text{ }i}^{j}(\mathbf{x},\mathbf{y})$ are the \emph{coefficients}
of the nonlinear connection $\mathcal{H}$. The basis $\left\{ \dfrac{\delta
}{\delta x^{i}},\dfrac{\partial }{\partial y^{j}}\right\} =\left\{ \delta
_{i},\dot{\partial}_{j}\right\} $ is called the \emph{adapted basis.}

The dual basis to the adapted basis is $\left\{ dx^{i},\delta y^{j}\right\} $%
, with
\begin{equation}
\delta y^{j}=dy^{j}+H_{\text{ }i}^{j}(\mathbf{x},\mathbf{y})dx^{i}.
\label{23}
\end{equation}

A \emph{distinguished tensor }(or \emph{d-tensor}) \emph{field of (r,s)-type}
is a quantity whose components transform like a tensor under the first
coordinate transformation (19) on $T\mathcal{M}$ (namely they change as
tensor in $\mathcal{M}$). For instance, for a d-tensor of type (1,2):
\begin{equation}
R_{\text{ \ }jk}^{\prime i}=\frac{\partial x^{\prime i}}{\partial x^{s}}%
\frac{\partial x^{r}}{\partial x^{\prime j}}\frac{\partial x^{p}}{\partial
x^{\prime k}}R_{\text{ }rp}^{s}.  \label{24}
\end{equation}%
In particular, both $\left\{ \dfrac{\delta }{\delta x^{i}}\right\} $ and $%
\left\{ \dfrac{\partial }{\partial y^{j}}\right\} $ are d-(covariant)
vectors, whereas $\left\{ dx^{i}\right\} $, $\left\{ \delta y^{j}\right\} $
are d-(contravariant) vectors.

A \emph{generalized Lagrange space} is a pair $\mathcal{GL}^{N}$=($\mathcal{M%
}$, $g_{ij}(\mathbf{x},\mathbf{y})$), with $g_{ij}(\mathbf{x},\mathbf{y})$
being a d-tensor of type (0,2) (covariant) on the manifold $T\mathcal{M}$ ,
which is symmetric, non-degenerate\footnote{%
Namely it must be $rank\left\Vert g_{ij}(\mathbf{x},\mathbf{y})\right\Vert
=N $.} and of constant signature.

A function
\begin{equation}
L:(\mathbf{x},\mathbf{y})\in T\mathcal{M}\rightarrow L(\mathbf{x},\mathbf{y}%
)\in \mathcal{R}  \label{25}
\end{equation}%
differentiable on $\widehat{T\mathcal{M}}$ and continuous on the null
section of $\pi $ is named a \emph{regular Lagrangian} if the Hessian of $L$
with respect to the variables $y^{i}$ is non-singular.

A generalized Lagrange space $\mathcal{GL}^{N}$=($\mathcal{M}$, $g_{ij}(%
\mathbf{x},\mathbf{y})$) is reducible to a \emph{Lagrange space} $\mathcal{L}%
^{N}$ if there is a regular Lagrangian $L$ satisfying
\begin{equation}
g_{ij}=\frac{1}{2}\frac{\partial ^{2}L}{\partial y^{i}\partial y^{j}}
\label{26}
\end{equation}%
on $\widehat{T\mathcal{M}}$ . In order that $\mathcal{GL}^{N}$ is reducible
to a Lagrange space, a necessary condition is the total symmetry of the
d-tensor $\dfrac{\partial g_{ij}}{\partial y^{k}}$. If such a condition is
satisfied, and $g_{ij}$ are 0-homogeneous in the variables $y^{i}$, then the
function $L=g_{ij}(\mathbf{x},\mathbf{y})y^{i}y^{j}$ is a solution of the
system (26). In this case, the pair ($\mathcal{M},L$) is a \emph{Finsler
space}\footnote{%
Let us recall that a Finsler space is a couple ($\mathcal{M}$, $\Phi $),
where $\mathcal{M}$ is be an N-dimensional differential manifold and $\Phi :T%
\mathcal{M}$ $\Rightarrow \mathcal{R}$ a function $\Phi (\mathbf{x,\xi )}$
defined for $\mathbf{x\in \mathcal{M}}$ and\textbf{\ }$\mathbf{\xi \in }$ $%
T_{\mathbf{x}}\mathcal{M}$ such that $\Phi (\mathbf{x,\cdot )}$ is a
possibly non symmetric norm on $T_{\mathbf{x}}\mathcal{M}$ .
\par
Notice that every Riemann manifold ($\mathcal{M}$, $\mathbf{g}$) is also a
Finsler space, the norm $\Phi (\mathbf{x,\xi )}$ being the norm induced by
the scalar product $\mathbf{g(x)}$.
\par
A finite-dimensional Banach space is another simple example of Finsler
space, where $\Phi (\mathbf{x,\xi )\equiv }\left\Vert \mathbf{\xi }%
\right\Vert $ .
\par
{}} ($\mathcal{M},\Phi $), with $\Phi ^{2}=L$. One says that $\mathcal{GL}%
^{N}$ is reducible to a Finsler space.

Of course, $\mathcal{GL}^{N}$ reduces to a pseudo-Riemannian (or Riemannian)
space ($\mathcal{M},g_{ij}(\mathbf{x})$) if the d-tensor $g_{ij}(\mathbf{x},%
\mathbf{y})$ does not depend on $\mathbf{y}$. On the contrary, if $g_{ij}(%
\mathbf{x},\mathbf{y})$ depends only on $\mathbf{y}$ (at least in preferred
charts), it is a generalized Lagrange space which is locally Minkowskian.

Since, in general,\ a generalized Lagrange space is not reducible to a
Lagrange one, it cannot be studied by means of the methods of symplectic
geometry, on which --- as is well known --- analytical mechanics is based.

A linear $\mathcal{H-}$connection on $T\mathcal{M}$ (or on $\widehat{T%
\mathcal{M}}$) is defined by a couple of geometrical objects $\mathcal{C}%
\Gamma (\mathcal{H})=(L_{\text{ }jk}^{i},C_{\text{ }jk}^{i})$ on $T\mathcal{M%
}$ with different transformation properties under the coordinate
transformation (19). Precisely, $L_{\text{ }jk}^{i}(\mathbf{x},\mathbf{y})$
transform like the coefficients of a linear connection on $\mathcal{M}$,
whereas $C_{\text{ }jk}^{i}(\mathbf{x},\mathbf{y})$ transform like a
d-tensor of type (1,2). $\mathcal{C}\Gamma (\mathcal{H})$ is called \emph{%
the metrical canonical }$\mathcal{H-}$\emph{connection} of the generalized
Lagrange space $\mathcal{GL}^{N}$.

In terms of $L_{\text{ }jk}^{i}$ and $C_{\text{ }jk}^{i}$ one can define two
kinds of covariant derivatives: a\emph{\ covariant horizontal (h-) derivative%
}, denoted by \textquotedblright $\shortmid $\textquotedblright , and a
\emph{covariant vertical (v-) derivative}, denoted by \textquotedblright $%
\mid $\textquotedblright . For instance, for the d-tensor $g_{ij}(\mathbf{x},%
\mathbf{y})$ one has
\begin{equation}
\left\{
\begin{array}{c}
g_{ij\shortmid k}=\dfrac{\delta g_{ij}}{\delta x^{k}}-g_{sj}L_{\text{ }%
ik}^{s}-g_{is}L_{\text{ }jk}^{s};\bigskip \\
g_{ij\mid k}=\dfrac{\partial g_{ij}}{\partial x^{k}}-g_{sj}C_{\text{ }%
ik}^{s}-g_{is}C_{\text{ }jk}^{s}.%
\end{array}%
\right.  \label{27}
\end{equation}%
The two derivatives $g_{ij\shortmid k}$ and $g_{ij\mid k}$ are both
d-tensors of type (0,3).

The coefficients of $\mathcal{C}\Gamma (\mathcal{H})$ can be expressed in
terms of the following \emph{generalized Christoffel symbols}:
\begin{equation}
\left\{
\begin{array}{c}
L_{\text{ }jk}^{i}=\frac{1}{2}g^{is}\left( \dfrac{\delta g_{sj}}{\delta x^{k}%
}+\dfrac{\delta g_{ks}}{\delta x^{j}}+\dfrac{\delta g_{jk}}{\delta x^{s}}%
\right) ;\bigskip \\
C_{\text{ }jk}^{i}=\frac{1}{2}g^{is}\left( \dfrac{\partial g_{sj}}{\partial
x^{k}}+\dfrac{\partial g_{ks}}{\partial x^{j}}+\dfrac{\partial g_{jk}}{%
\partial x^{s}}\right) .%
\end{array}%
\right.  \label{28}
\end{equation}

\paragraph{Curvature and torsion in a generalized Lagrange space}

By means of the connection $\mathcal{C}\Gamma (\mathcal{H})$ it is possible
to define a \emph{d-curvature} in $T\mathcal{M}$ by means of the tensors $%
R_{j\text{ }kh}^{\text{ }i}$, $S_{j\text{ }kh}^{\text{ }i}$ and $P_{j\text{ }%
kh}^{\text{ }i}$ given by
\begin{eqnarray}
R_{j\text{ }kh}^{\text{ }i} &=&\frac{\delta L_{\text{ }jk}^{i}}{\delta x^{h}}%
-\frac{\delta L_{\text{ }jh}^{i}}{\delta x^{k}}+L_{\text{ }jk}^{r}L_{\text{ }%
rh}^{i}-L_{\text{ }jh}^{r}L_{\text{ }rk}^{i}+C_{\text{ }jr}^{i}R_{\text{ }%
kh}^{r};\bigskip  \notag \\
S_{j\text{ }kh}^{\text{ }i} &=&\frac{\partial C_{\text{ }jk}^{i}}{\partial
y^{h}}-\frac{\partial C_{\text{ }jh}^{i}}{\partial y^{k}}+C_{\text{ }%
jk}^{r}C_{\text{ }rh}^{i}-C_{\text{ }jh}^{r}C_{\text{ }rk}^{i};\bigskip
\notag \\
P_{j\text{ }kh}^{\text{ }i} &=&\frac{\partial L_{\text{ }jk}^{i}}{\partial
y^{h}}-C_{\text{ }j\shortmid h}^{i}+C_{\text{ }jr}^{i}P_{\text{ }kh}^{r}.
\label{30}
\end{eqnarray}%
Here, the d-tensor $R_{jk}^{i}$ is related to the bracket of the basis $%
\left\{ \dfrac{\delta }{\delta x^{i}}\right\} $ :
\begin{equation}
\left[ \dfrac{\delta }{\delta x^{i}},\dfrac{\delta }{\delta xj}\right] =R_{%
\text{ }ij}^{s}\frac{\partial }{\partial y^{s}}  \label{31}
\end{equation}%
and is explicitly given by\footnote{$R_{jk}^{i}$ plays the role of a
curvature tensor of the nonlinear connection $\mathcal{H}$. The
corresponding tensor of torsion is instead
\begin{equation*}
t_{jk}^{i}=\dfrac{\partial H_{j}^{i}}{\partial y^{k}}-\dfrac{\partial
H_{k}^{i}}{\partial y^{j}}.
\end{equation*}%
}
\begin{equation}
R_{\text{ }jk}^{i}=\dfrac{\delta H_{\text{ }j}^{i}}{\delta x^{k}}-\dfrac{%
\delta H_{\text{ }k}^{i}}{\delta x^{j}}.  \label{32}
\end{equation}%
The tensor $P_{\text{ }jk}^{i}$, together with $T_{\text{ }jk}^{i}$, $S_{%
\text{ }jk}^{i}$, defined by
\begin{eqnarray}
P_{\text{ }jk}^{i} &=&\frac{\partial H_{\text{ }j}^{i}}{\partial y^{k}}-L_{%
\text{ }jk}^{i};\bigskip  \notag \\
T_{\text{ }jk}^{i} &=&L_{\text{ }jk}^{i}-L_{\text{ }kj}^{i};\bigskip  \notag
\\
S_{\text{ }jk}^{i} &=&C_{\text{ }jk}^{i}-C_{\text{ }kj}^{i}  \label{33}
\end{eqnarray}%
are \emph{the d-tensors of torsion of the metrical connection} $\mathcal{C}%
\Gamma (\mathcal{H})$.

>From the curvature tensors one can get the corresponding Ricci tensors of $%
\mathcal{C}\Gamma (\mathcal{H})$:
\begin{equation}
\left\{
\begin{array}{cc}
R_{ij}=R_{i\text{ }js}^{\text{ }s};\bigskip & S_{ij}=S_{i\text{ }js}^{\text{
}s};\bigskip \\
\overset{1}{P}_{ij}=P_{i\text{ }js}^{\text{ }s} & \overset{2}{P}_{ij}=P_{i%
\text{ }sj}^{\text{ }s},%
\end{array}%
\right.  \label{34}
\end{equation}%
and the scalar curvatures
\begin{equation}
\begin{array}{cc}
R=g^{ij}R_{ij}; & S=g^{ij}S_{ij}.%
\end{array}
\label{35}
\end{equation}

Finally,\emph{\ the deflection d-tensors associated to the connection} $%
\mathcal{C}\Gamma (\mathcal{H})$ are
\begin{equation}
\left\{
\begin{array}{c}
D_{\text{ }j}^{i}=y_{\shortmid j}^{i}=-H_{\text{ }j}^{i}+y^{s}L_{\text{ }%
sj}^{i};\bigskip \\
d_{\text{ }j}^{i}=y_{\mid j}^{i}=\delta _{\text{ }j}^{i}+y^{s}C_{sj}^{i},%
\end{array}%
\right.  \label{36}
\end{equation}%
namely the h- and v-covariant derivatives of the Liouville vector fields.

In the generalized Lagrange space $\mathcal{GL}^{N}$ it is possible to write
the Einstein equations with respect to the canonical connection $\mathcal{C}%
\Gamma (\mathcal{H})$ as follows:
\begin{equation}
\left\{
\begin{array}{cc}
R_{ij}-\frac{1}{2}Rg_{ij}=\kappa \overset{H}{T}_{ij};\bigskip & \overset{1}{P%
}_{ij}=\kappa \overset{1}{T}_{ij};\bigskip \\
S_{ij}-\frac{1}{2}Sg_{ij}=\kappa \overset{V}{T}_{ij}; & \overset{2}{P}%
_{ij}=\kappa \overset{2}{T}_{ij},%
\end{array}%
\right.  \label{37}
\end{equation}%
where $\kappa $ is a constant and $\overset{H}{T}_{ij}$, $\overset{V}{T}%
_{ij} $, $\overset{1}{T}_{ij}$, $\overset{2}{T}_{ij}$ are the components of
the energy-momentum tensor.

\subsubsection{Generalized Lagrangian Structure of $\widetilde{M}$}

On the basis of the previous considerations, let us analyze the geometrical
structure of the deformed Minkowski space of DSR $\widetilde{M}$, endowed
with the by now familiar metric $g_{\mu \nu ,DSR}(E).$ As said in Sect.2, $E$
is the energy of the process measured by the detectors in Minkowskian
conditions. Therefore, $E$ is a function of the velocity components, $u^{\mu
}=dx^{\mu }/d\tau $, where $\tau $ is the (Minkowskian) proper time\footnote{%
Contrarily to ref.\cite{jan}, we shall not consider the restrictive case of
a classical (non-relativistic) expression of the energy, but assume a
general dependence of $E$ on the velocity (eq.(38)).}:
\begin{equation}
E=E\left( \frac{dx^{\mu }}{d\tau }\right) .  \label{38}
\end{equation}

The derivatives $dx^{\mu }/d\tau $ define a contravariant vector tangent to $%
M$ at $x$, namely they belong to $TM_{\mathbf{x}}$. We shall denote this
vector (according to the notation of the previous Subsubsection) by $\mathbf{%
y}=(y^{\mu })$. Then, ($\mathbf{x}$,$\mathbf{y}$) is a point of the tangent
bundle to $M$. \ We can therefore consider the generalized Lagrange space $%
\mathcal{GL}^{4}=(M,g_{\mu \nu }(\mathbf{x}$,$\mathbf{y))}$, with
\begin{equation}
\left\{
\begin{array}{c}
g_{\mu \nu }(\mathbf{x},\mathbf{y)=}g_{\mu \nu DSR}(E(\mathbf{x},\mathbf{y)),%
} \\
\\
E(\mathbf{x},\mathbf{y)=}E(\mathbf{y).}%
\end{array}%
\right.  \label{39}
\end{equation}

Then, it is possible to prove the following theorem \cite{jan}:

\emph{The pair} $\mathcal{GL}^{4}=(M,g_{DSR,\mu \nu }(\mathbf{x}$,$\mathbf{%
y))}\equiv \widetilde{M}$\emph{\ is a generalized Lagrange space which is
not reducible to a Riemann space, or to a Finsler space, or to a Lagrange
space.}

Notice that such a result is strictly related to the fact that the deformed
metric tensor of DSR is diagonal.

If an external electromagnetic field $F_{\mu \nu }$ is present in the
Minkowski space $M$, in $\widetilde{M}$ the deformed electromagnetic field
is given by $\widetilde{F}_{\text{ }\nu }^{\mu }(\mathbf{x},\mathbf{y)}%
=g_{DSR}^{\mu \rho }F_{\rho \nu }(\mathbf{x)}$ (see Eq.(8)). Such a field is
a d-tensor and is called \emph{the electromagnetic tensor of the generalized
Lagrange space}. Then, the nonlinear connection $\mathcal{H}$ is given by
\begin{equation}
H_{\text{ }\nu }^{\mu }=\left\{
\begin{array}{c}
\mu \\
\nu \rho%
\end{array}%
\right\} y^{\rho }-\widetilde{F}_{\text{ }\nu }^{\mu }(\mathbf{x},\mathbf{y),%
}  \label{40}
\end{equation}%
where $\left\{
\begin{array}{c}
\mu \\
\nu \rho%
\end{array}%
\right\} $, the Christoffel symbols of the Minkowski metric $g_{\mu \nu }$,
are zero, so that
\begin{equation}
H_{\text{ }\nu }^{\mu }=-\widetilde{F}_{\text{ }\nu }^{\mu }(\mathbf{x},%
\mathbf{y),}  \label{41}
\end{equation}%
namely, \emph{the connection coincides with the deformed field.}

The adapted basis of the distribution $\mathcal{H}$ reads therefore
\begin{equation}
\frac{\delta }{\delta x^{\mu }}=\frac{\partial }{\partial x^{\mu }}+%
\widetilde{F}_{\text{ }\mu }^{\nu }(\mathbf{x},\mathbf{y)}\frac{\partial }{%
\partial y^{\nu }}.  \label{42}
\end{equation}%
The local covector field of the dual basis (cfr. Eq.(23)) is given by
\begin{equation}
\delta y^{\mu }=dy^{\mu }-\widetilde{F}_{\text{ }\nu }^{\mu }(\mathbf{x},%
\mathbf{y})dx^{\nu }.  \label{43}
\end{equation}

\subsubsection{Canonical Metric Connection of $\widetilde{M}$}

The derivation operators applied to the deformed metric tensor of the space $%
\mathcal{GL}^{4}=\widetilde{M}$ yield
\begin{equation}
\frac{\delta g_{DSR\mu \nu }}{\delta x^{\rho }}=\frac{\partial g_{DSR\mu \nu
}}{\partial x^{\rho }}+\widetilde{F}_{\rho }^{\sigma }\frac{\partial
g_{DSR\mu \nu }}{\partial y^{\sigma }}=\widetilde{F}_{\text{ }\rho }^{\sigma
}\frac{\partial g_{DSR\mu \nu }}{\partial E}\frac{\partial E}{\partial
y^{\sigma }},  \label{44}
\end{equation}%
\begin{equation}
\frac{\partial g_{DSR\mu \nu }}{\partial y^{\sigma }}=\frac{\partial
g_{DSR\mu \nu }}{\partial E}\frac{\partial E}{\partial y^{\sigma }}.
\label{45}
\end{equation}%
Then, the coefficients of the canonical metric connection $\mathcal{C}\Gamma
(\mathcal{H})$ in $\widetilde{M}$ (see Eq.(28)) are given by
\begin{equation}
\left\{
\begin{array}{c}
L_{\text{ }\nu \rho }^{\mu }=\frac{1}{2}g_{DSR}^{\mu \sigma }\dfrac{\partial
E}{\partial y^{\alpha }}\left( \dfrac{\partial g_{DSR\sigma \nu }}{\partial E%
}\widetilde{F}_{\text{ }\rho }^{\alpha }+\dfrac{\partial g_{DSR\sigma \rho }%
}{\partial E}\widetilde{F}_{\text{ }\nu }^{\alpha }-\dfrac{\partial
g_{DSR\nu \rho }}{\partial E}\widetilde{F}_{\text{ }\sigma }^{\alpha
}\right) , \\
\\
C_{\text{ }\nu \rho }^{\mu }=\frac{1}{2}g_{DSR}^{\mu \sigma }\dfrac{\partial
E}{\partial y^{\alpha }}\left( \dfrac{\partial g_{DSR\sigma \nu }}{\partial E%
}\delta _{\text{ }\rho }^{\alpha }+\dfrac{\partial g_{DSR\sigma \rho }}{%
\partial E}\delta _{\text{ }\nu }^{\alpha }-\dfrac{\partial g_{DSR\nu \rho }%
}{\partial E}\delta _{\text{ }\sigma }^{\alpha }\right) .%
\end{array}%
\right.  \label{46}
\end{equation}

The vanishing of the electromagnetic field tensor, $F_{\text{ }\rho
}^{\alpha }=0$, implies $L_{\text{ }\nu \rho }^{\mu }=0$.

One can define the deflection tensors associated to the metric connection $%
\mathcal{C}\Gamma (\mathcal{H})$ as follows (cfr. Eq.(36)):
\begin{eqnarray}
D_{\text{ }\nu }^{\mu } &=&y_{\text{ }\shortmid \nu }^{\mu }=\frac{\delta
y^{\mu }}{\delta x^{\nu }}+y^{\alpha }L_{\text{ }\alpha \nu }^{\mu }=%
\widetilde{F_{\text{ }\nu }^{\mu }}+y^{\alpha }L_{\text{ }\alpha \nu }^{\mu
};\bigskip  \notag \\
d_{\text{ }\nu }^{\mu } &=&y_{\text{ }\mid \nu }^{\mu }=\delta _{\text{ }\nu
}^{\mu }+y^{\alpha }C_{\text{ }\alpha \nu }^{\mu }.  \label{47}
\end{eqnarray}%
The covariant components of these tensors read
\begin{gather}
D_{\mu \nu }=g_{\mu \sigma ,DSR}D_{\text{ }\nu }^{\sigma }=g_{\mu \sigma
,DSR}\left( \widetilde{F_{\text{ }\nu }^{\sigma }}+y^{\alpha }L_{\text{ }%
\alpha \nu }^{\sigma }\right) =\bigskip  \notag \\
=F_{\mu \nu }(\mathbf{x})+\frac{1}{2}y^{\sigma }\dfrac{\partial E}{\partial
y^{\alpha }}\left( \dfrac{\partial g_{DSR\mu \sigma }}{\partial E}\widetilde{%
F}_{\text{ }\nu }^{\alpha }+\dfrac{\partial g_{DSR\mu \nu }}{\partial E}%
\widetilde{F}_{\text{ }\sigma }^{\alpha }-\dfrac{\partial g_{DSR\sigma \nu }%
}{\partial E}\widetilde{F}_{\text{ }\mu }^{\alpha }\right) ;  \notag \\
\notag \\
d_{\mu \nu }=g_{\mu \sigma ,DSR}d_{\text{ }\nu }^{\sigma }=\bigskip  \notag
\\
=g_{DSR,\mu \nu }+\frac{1}{2}y^{\sigma }\dfrac{\partial E}{\partial
y^{\alpha }}\left( \dfrac{\partial g_{DSR\mu \sigma }}{\partial E}\delta _{%
\text{ }\nu }^{\alpha }+\dfrac{\partial g_{DSR\mu \nu }}{\partial E}\delta _{%
\text{ }\sigma }^{\alpha }-\dfrac{\partial g_{DSR\sigma \nu }}{\partial E}%
\delta _{\text{ }\mu }^{\alpha }\right) .  \label{48}
\end{gather}

It is important to stress explicitly that, on the basis of the results of
3.2.1, \emph{the deformed Minkowski space }$\widetilde{M}$ \emph{does
possess curvature and torsion,} namely it is endowed with a very rich
geometrical structure. This permits to understand the variety of new
physical phenomena that occur in it (as compared to the standard Minkowski
space) \cite{carmig1,carmig2}.

Following ref.\cite{jan}, let us show how the formalism of the generalized
Lagrange space allows one to recover some results on the phenomenological
energy-dependent metrics discussed in Sect.2.

Consider the following metric ($c=1$):
\begin{equation}
ds^{2}=a(E)dt^{2}+(dx^{2}+dy^{2}+dz^{2})  \label{49}
\end{equation}%
where $a(E)$ is an arbitrary function of the energy and spatial isotropy ($%
b^{2}=1$) has been assumed. In absence of an external electromagnetic field (%
$F_{\mu \nu }=0$), the non-vanishing components $C_{\text{ }\nu \rho }^{\mu
} $ of the canonical metric connection $\mathcal{C}\Gamma (\mathcal{H})$
(see Eq.(46)) are
\begin{equation}
\left\{
\begin{array}{cccc}
C_{00}^{0}=\dfrac{a^{\prime }}{a}y^{0},\bigskip & C_{01}^{0}=-\dfrac{%
a^{\prime }}{a}y^{1},\bigskip & C_{02}^{0}=-\dfrac{a^{\prime }}{a}%
y^{2},\bigskip & C_{03}^{0}=\dfrac{a^{\prime }}{a}y^{3},\bigskip \\
C_{00}^{1}=-a^{\prime }y^{1}, & C_{00}^{2}=-a^{\prime }y^{2}, &
C_{00}^{0}=-a^{\prime }y^{3}, &
\end{array}%
\right.  \label{50}
\end{equation}%
where the prime denotes derivative with respect to $E$: $a^{\prime }=\dfrac{%
da}{dE}.$

According to the formalism of generalized Lagrange spaces, we can write the
Einstein equations in vacuum corresponding to the metrical connection of the
deformed Minkowski space (see Eqs.(37)). It is easy to see that the
independent equations are given by
\begin{gather}
a^{\prime }=0;\bigskip  \label{51} \\
2aa^{\prime \prime }-\left( a^{\prime }\right) ^{2}=0.  \label{52}
\end{gather}%
The first equation has the solution $a=const.$, namely we get the Minkowski
metric. Eq.(52) has the solution
\begin{equation}
a(E)=\frac{1}{4}\left( a_{0}+\frac{E}{E_{0}}\right) ^{2},  \label{53}
\end{equation}%
where $a_{0}$ and $E_{0}$ are two integration constants.

This solution represents the time coefficient of an over-Minkowskian metric.
For $a_{0}=0$ it coincides with (the time coefficient of) the
phenomenological metric of the strong interaction, Eq.(6). On the other
hand, by choosing $a_{0}=1$, one gets the time coefficient of the metric for
gravitational interaction, Eq.(6').

In other words, \emph{considering }$\widetilde{M}$\emph{\ as a generalized
Lagrange space permits to recover (at least partially) the metrics of two
interactions (strong and gravitational) derived on a phenomenological basis.
}

It is also worth noticing that this result shows that \emph{a spacetime
deformation (of over-Minkowskian type) exists even in absence of an external
electromagnetic field }(remember that Eqs.(51),(52) have been derived by
assuming $F_{\mu \nu }=0$).

\subsubsection{Intrinsic Physical Structure of a Deformed Minkowski Space:
Gauge Fields}

As we have seen, the deformed Minkowski space $\widetilde{M}$ , considered
as a generalized Lagrange space, is endowed with a rich geometrical
structure. But the important point, to our purposes, is the presence of a
physical richness, intrinsic to $\widetilde{M}$ . Indeed, let us introduce
the following \emph{internal electromagnetic field tensors }on $\mathcal{GL}%
^{4}=\widetilde{M}$, defined in terms of the deflection tensors\emph{:}
\begin{gather}
\mathcal{F}_{\mu \nu }\equiv \frac{1}{2}\left( D_{\mu \nu }-D_{\nu \mu
}\right) =\bigskip  \notag \\
=F_{\mu \nu }(\mathbf{x})+\frac{1}{2}y^{\sigma }\dfrac{\partial E}{\partial
y^{\alpha }}\left( \dfrac{\partial g_{DSR\mu \sigma }}{\partial E}\widetilde{%
F}_{\text{ }\nu }^{\alpha }-\dfrac{\partial g_{DSR\nu \sigma }}{\partial E}%
\widetilde{F}_{\text{ }\mu }^{\alpha }\right)  \label{54}
\end{gather}%
(\emph{horizontal electromagnetic internal tensor}) and
\begin{gather}
f_{\mu \nu }\equiv \frac{1}{2}\left( d_{\mu \nu }-d_{\nu \mu }\right)
=\bigskip  \notag \\
=\frac{1}{2}y^{\sigma }\dfrac{\partial E}{\partial y^{\alpha }}\left( \dfrac{%
\partial g_{DSR\mu \sigma }}{\partial E}\delta _{\text{ }\nu }^{\alpha }-%
\dfrac{\partial g_{DSR\nu \sigma }}{\partial E}\delta _{\text{ }\mu
}^{\alpha }\right)  \label{55}
\end{gather}%
(\emph{vertical electromagnetic internal tensor}).

The internal electromagnetic h- and v-fields $\mathcal{F}_{\mu \nu }$ and $%
f_{\mu \nu }$ satisfy the following \emph{generalized Maxwell equations}
\begin{eqnarray}
2\left( \mathcal{F}_{\mu \nu \shortmid \rho }+\mathcal{F}_{\nu \rho
\shortmid \mu }+\mathcal{F}_{\rho \mu \shortmid \nu }\right) &=&y^{\alpha
}\left( R_{\text{ }\mu \nu }^{\beta }C_{\beta \alpha \rho }+R_{\text{ }\nu
\rho }^{\beta }C_{\beta \alpha \mu }+R_{\text{ }\rho \mu }^{\beta }C_{\beta
\alpha \nu }\right) ,\bigskip  \notag \\
R_{\text{ }\mu \nu }^{\beta } &=&g^{\beta \sigma }\frac{\partial F_{\mu \nu }%
}{\partial x^{\sigma }};  \label{56}
\end{eqnarray}%
\begin{equation}
\mathcal{F}_{\mu \nu \mid \rho }+\mathcal{F}_{\nu \rho \mid \mu }+\mathcal{F}%
_{\rho \mu \mid \nu }=f_{\mu \nu \shortmid \rho }+f_{\nu \rho \shortmid \mu
}+f_{\rho \mu \shortmid \nu };  \label{57}
\end{equation}%
\begin{equation}
f_{\mu \nu \mid \rho }+f_{\nu \rho \mid \mu }+f_{\rho \mu \mid \nu }=0.
\label{58}
\end{equation}

Let us stress explicitly the different nature of the two internal
electromagnetic fields. In fact, the horizontal field $\mathcal{F}_{\mu \nu
} $ is strictly related to the presence of the external electromagnetic
field $F_{\mu \nu }$ , and vanishes if $F_{\mu \nu }=0$. On the contrary,%
\emph{\ the vertical field }$f_{\mu \nu }$\emph{\ has a geometrical origin,
and depends only on the deformed metric tensor }$g_{DSR\mu \nu }(E(\mathbf{y}%
))$\emph{\ of }$\mathcal{GL}^{4}=\widetilde{M}$\emph{\ and on }$E(\mathbf{y}%
) $\emph{. Therefore, it is present also in space-time regions where no
external electromagnetic field occurs.} As we shall see, this fact has deep
physical implications.

A few remarks are in order. First, the main results obtained for the
(abelian) electromagnetic field can be probably generalized (with suitable
changes) to non-abelian gauge fields. Second, the presence of the internal
electromagnetic h- and v-fields $\mathcal{F}_{\mu \nu }$ and $f_{\mu \nu }$,
intrinsic to the geometrical structure of $\widetilde{M}$ as a generalized
Lagrange space, is the cornerstone to build up a \emph{dynamics (of merely
geometrical origin) internal to the deformed Minkowski space.}

The important point worth emphasizing is that \emph{such an intrinsic
dynamics springs from gauge fields.} Indeed, the two internal fields $%
\mathcal{F}_{\mu \nu }$ and $f_{\mu \nu }$ \ (in particular the latter one)
do satisfy equations of the gauge type (cfr. Eqs.(57)-(58)). Then, we can
conclude that \emph{the (energy-dependent) deformation of the metric of }$%
\widetilde{M}$\emph{\ , which induces its geometrical structure as
generalized Lagrange space, leads in turn to the appearance of (internal)
gauge fields}.

Such a fundamental result can be schematized as follows:
\begin{equation}
\widetilde{M}=\left( M,g_{DSR\mu \nu }(E)\right) \Longrightarrow \mathcal{GL}%
^{4}=(M,g_{\mu \nu }(\mathbf{x},\mathbf{y))\Longrightarrow }\left(
\widetilde{M},\mathcal{F}_{\mu \nu },f_{\mu \nu }\right)  \label{59}
\end{equation}%
(with self-explanatory meaning of the notation).

We want also to stress explicitly that this result follows by the fact that,
in deforming the metric of the space-time, \emph{we assumed the energy as
the physical (non-metric) observable on which letting the metric
coefficients depend} . This is crucial in stating the generalized Lagrangian
structure of $\widetilde{M}$, as shown above.

\subsection{Possible evidence for DSR internal gauge fields: Shadow of light}

We want now to discuss some results on anomalous interference effects, which
admit a quite straightforward interpretation in terms of the intrinsic gauge
fields of DSR.

\begin{figure}[tbp]
\begin{center}
\includegraphics[width=0.8\textwidth]{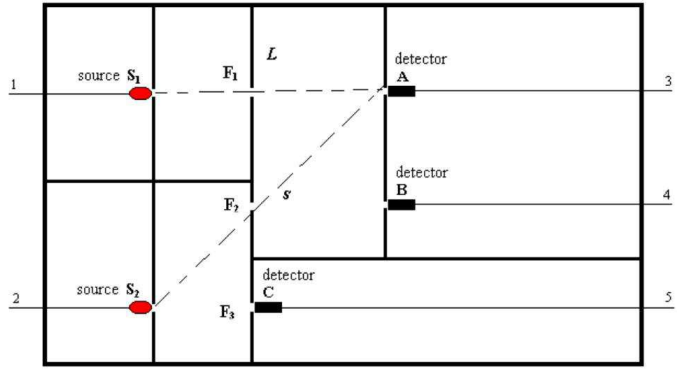}
\end{center}
\caption{Schematic layout of the box used to detect the anomalous
interference effect.} \label{exp_scheme}
\end{figure}

In double-slit-like experiments in the infrared range, we collected
evidences of an anomalous behaviour of photon systems under
particular (energy and space) constraints
\cite{shad1,shad2,shad3,shad4}. The experimental set-up is reported
in Fig.\ref{exp_scheme}.


This layout shows the horizontal view of the interior of a closed box
divided into different rooms by panels. The box was 20 $cm$ long, 12 $cm$
large and 7 $cm$ high. It contained two infrared LEDs S$_{1}$ and S$_{2}$,
three detectors A, B and C (either photodiodes or phototransistors) and
three apertures F$_{1}$, F$_{2}$ and F$_{3}$. The source S$_{1}$ was aligned
with the detector A through the aperture F$_{1}$, the source S$_{2}$ was
aligned with the detector C which was right on the aperture F$_{3}$. The
detector B was in front of the aperture F$_{2}$ and did not receive any
photon directly. The position of the detectors, the sources and the
apertures was designed so that the detector A was not influenced by the
lighting state of the source S$_{2}$ according to the laws of physics
governing photons propagation. In other words, A did not have to distinguish
whether S$_{2}$ was on or off. Besides, in order to prevent reflections of
photons, the internal surfaces of the box had been coated by an absorbing
material. While the detectors B and C were controlling detectors, A was
devoted to perform the actual experiment. In particular, we compared the
signal, measured on A when S$_{1}$ was on and S$_{2}$ was off, with the
signal on A when both sources S$_{1}$ and S$_{2}$ were on. As to what it has
been said about the incapability of A to distinguish between S$_{2}$ off or
on, these two compared conditions were expected to produce compatible
results. However, it turned out that the sampling of the signal on A with S$%
_{1}$ on and S$_{2}$ on and the sampling of the signal on A when only S$_{1}$
was on do not belong to the same population and are represented by two
different gaussian distributions whose mean values are significantly
different. Besides, the difference between the two mean values was less than
4.5 $\mu eV$ , as predicted by the theory of Deformed Space-time \cite%
{carmig1,carmig2}. Since it was experimentally verified that no photons
passed through the aperture F$_{2}$, this result shows an anomalous
behaviour of the photon system. The same experiment was carried out by
different sources, detectors, by two different boxes and different measuring
systems. Nevertheless, every time we obtained the same anomalous result \cite%
{shad2,shad3,shad4}. Moreover, the same kind of geometrical structure and
the same spatial distances were used in other kind of experiments carried
out in the microwave region of the spectrum and by a laser system \cite%
{ranf1,ranf2,carmig1,carmig2}. Although these experiments had completely
different experimental set-ups from our initial one, they succeeded in
finding out the same kind of anomalous behaviour that we had found out by
the box experiments.

The anomalous effect in photon systems, at least in those experimental
set-ups that were used, disagrees both with standard quantum mechanics
(Copenhagen interpretation) and with classical and quantum electrodynamics.
Some possible interpretations can be given in terms of either the existence
of de Broglie--Bohm pilot waves associated to photons, and/or the breakdown
of local Lorentz invariance (LLI) \cite{shad1,shad2,shad3,shad4}. Besides,
it turns out that it is also possible to move a step forward and hypothesise
the existence of an intriguing connection between the pilot wave
interpretation and that involving LLI breakdown. One might assume that the
pilot wave is, in the framework of LLI breakdown, a local deformation of the
flat Minkowskian spacetime.

The interpretation in terms of DSR is quite straightforward. Under the
energy threshold $E_{0,em}$=4.5 $\mu eV$, the metric of the electromagnetic
interaction is no longer Minkowskian. The corresponding space-time is
deformed. Such a space-time deformation shows up as the hollow wave
accompanying the photon, and is able to affect the motion of other photons.
This is the origin of the anomalous interference observed (\emph{shadow of
light}). The difference of signal measured by the detector A in all the
double-slit experiments can be regarded as the energy spent to deforme
space-time. In space regions where the external electromagnetic field is
present (regions of "standard" photon behavior), we can associate such
energy to the difference $\Delta \mathcal{E}$, Eq.(16), between the energy
density corresponding to the external e.m. field $F_{\mu \nu }$ and that of
the deformed one $\widetilde{F}_{\mu \nu }$ given by Eq.(8).

But it is known from the experimental results that the anomalous
interference effects observed can be explained in terms of the shadow of
light, namely in terms of the hollow waves present in space regions where no
external e.m. field occurs. How to account for this anomalous photon
behavior within DSR? The answer is provided by the internal structure of the
deformed Minkowski space discussed above. In fact, we have seen that the
structure of the deformed Minkowski space $\widetilde{M}$ as Generalized
Lagrange Space implies the presence of two internal e.m. fields, the
horizontal field $\mathcal{F}_{\mu \nu }$ and the vertical one, $f_{\mu \nu
} $. Whereas $\mathcal{F}_{\mu \nu }$ is strictly related to the presence of
the external electromagnetic field $F_{\mu \nu }$, vanishing if $F_{\mu \nu
}=0$, \emph{\ }the vertical field $f_{\mu \nu }$\ is geometrical in nature,
depending only on the deformed metric tensor $g_{DSR},_{\mu \nu }(E)$\ of $%
GL^{4}=\widetilde{M}$\ and on $E$. Therefore, it is present also in
space-time regions where no external electromagnetic field occurs. In our
opinion, the arising of the internal electromagnetic fields associated to
the deformed metric of $\widetilde{M}$ as Generalized Lagrange space is at
the very physical, \emph{dynamic }interpretation\emph{\ }of the experimental
results on the anomalous photon behavior. Namely, \emph{the dynamic effects
of the hollow wave of the photon, associated to the deformation of
space-time }--- which manifest themselves in the photon behavior
contradicting both classical and quantum electrodynamics ---\emph{, arise
from the presence of the internal v-electromagnetic field } $f_{\mu \nu }$
(in turn strictly connected to the geometrical structure of $\widetilde{M}$).

Moreover, as is well known, in relativistic theories, the vacuum is nothing
but Minkowski geometry. A LLI breaking connected to a deformation of the
Minkowski space is therefore associated to a lack of Lorentz invariance of
the vacuum. Then, the view by Kostelecky \cite{kost} that the breakdown of
LLI is related to the lack of Lorentz symmetry of the vacuum accords with
our results in the framework of DSR, provided that the quantum vacuum is
replaced by the geometric vacuum.

\section{Conclusions and perspectives}

As is well known, successfully embodying gauge fields in a space-time
structure is one of the basic goals of the research in theoretical physics
starting from the beginning of the XX century. The almost unique tool to
achieve such objective is increasing the number of space-time dimensions. In
such a kind of theories (whose prototype is the celebrated Kaluza-Klein
formalism), one preserves the usual (special-relativistic or
general-relativistic) structure of the four-dimensional space-time, and gets
rid of the non-observable extra dimensions by compactifying them (for
example to circles). Then the motions of the extra metric components over
the standard Minkowski space satisfy identical equations to gauge fields.
The gauge invariance of these fields is simply a consequence of the Lorentz
invariance in the enlarged space. In this framework, gauge fields are \emph{%
external} to the space-time, because they are \emph{added} to it by the
hypothesis of extra dimensions.

In the case of the DSR theory, gauge fields arise from the very geometrical,
basic structure of $\widetilde{M}$, namely they are a consequence of the
metric deformation. The arising gauge fields are \emph{intrinsic and
internal to the deformed space-time},\emph{\ and do not need to be added
from the outside.} As a matter of fact, \emph{DSR is the first theory based
on a four-dimensional space-time able to embody gauge fields in a natural
way. }

Such a conventional, intrinsic gauge structure is related to a \emph{given }%
deformed Minkowski space $\overline{\widetilde{M}}$, in which the deformed
metric is fixed:
\begin{equation}
\overline{\widetilde{M}}=\left( M,\bar{g}_{DSR\mu \nu }(E)\right) .
\label{60}
\end{equation}%
On the contrary, with varying $g_{DSR}$, we have another gauge-like
structure --- as already stressed in Sect.3 --- namely what we called an
external metric gauge. In the latter case, the gauge freedom amounts to
choosing the metric according to the interaction considered.

The circumstance that the deformed Minkowski space $\widetilde{M}$ is
endowed with the geometry of a generalized Lagrange space testifies the
richness of non-trivial mathematical properties present in the seemingly so
simple structure of the deformation of the Minkowski metric. In this
connection, let us recall that $\widetilde{M}$ (contrarily to the usual
Minkowski space) is not flat, but does possess curvature and torsion (see
3.2.1).

Let us stress that --- as already mentioned --- the deformed Minkowski space
$\widetilde{M}$ can be naturally embedded in a five-dimensional Riemannian
space $\mathcal{\Re }_{5}$ (see \cite{carmig1,carmig2}). We denoted by DR5
the generalized theory based on this five-dimensional space.

In embedding the deformed Minkowski space $\widetilde{M}$ in $\mathcal{\Re }%
_{5}$, \emph{energy does lose its character of dynamic parameter} (the role
it plays in DSR), \emph{by taking instead that of a true metrical
coordinate, }$E=x^{5}$, on the same footing of the space-time ones. This has
a number of basic implications. In such a change of role of energy, with the
consequent passage from $\widetilde{M}$ to $\mathcal{\Re }_{5}$, some of the
geometrical and dynamic features of DSR are lost, whereas others are still
present and new properties appear. The first one is of geometrical nature,
and is just the passage from a (flat) pseudoeuclidean metric to a genuine
(curved) Riemannian one. The other consequences pertain to both symmetries
and dynamics. Among the former, we recall the basic one --- valid at the
slicing level $x^{5}=const.$ ($dx^{5}=0$) ---, related to the Generalized
Lagrange Space structure of $\widetilde{M}$, which implies \emph{the natural
arising of gauge fields}, intimately related to the inner geometry of the
deformed Minkowski space. Let us also stress that, in the framework of $%
\mathcal{\Re }_{5}$, the dependence of the metric coefficients on a true
metric coordinate make them fully analogous to the gauge functions of
non-abelian gauge theories, thus implementing DR5 as a metric gauge theory
(in the sense specified in Subsect.3.1). Let us recall that the metric
homomorphisms of $\Re _{5}$ are strictly connected to the invariance under
what we called the Metric Gaugement Process of DSR (see Subsect.3.1).

Concerning the influence of the extra dimension on the physics in the
four-dimensional deformed space-time, points worth investigating are the
possible connection between Lorentz invariance in DR5 and the usual gauge
invariance, and the occurrence of parity violation as consequence of space
anisotropy when viewed from the standpoint of the space-time-energy manifold
$\Re _{5}$.

A further basic topic deserving study in DSR is the extension to the
non-abelian case of the results obtained for the abelian gauge fields (like
the e.m. one), based on the structure of the deformed Minkowski space $%
\widetilde{M}$ as Generalized Lagrange Space (see Subsubsect.3.2.1). In
other words, it would be worth verifying if also non-abelian internal gauge
fields can exist in absence of external fields, due to the intrinsic
geometry of $\widetilde{M}$.

\end{document}